-------------------------------------------------------------------
\documentstyle[12pt]{article}
\begin{document}

\newcommand{\regpreprint}[5]
{
\noindent
\begin{minipage}[t]{\textwidth}
\begin{center}
\framebox[\textwidth]{$\rule[6mm]{0mm}{0mm}$
\raisebox{1.3mm}{#1 Institut f\"ur Theoretische Physik der
Universit\"at Regensburg}}

\vspace{2mm}    \rule{\textwidth}{0.2mm}\\
\vspace{-4mm}   \rule{\textwidth}{1pt}
\mbox{ }    #2    \hfill    #3   \mbox{ }\\
\vspace{-2mm}   \rule{\textwidth}{1pt}\\
\vspace{-4.2mm} \rule{\textwidth}{0.2mm}
\end{center}
\end{minipage}

\vspace{15mm}

\begin{center} {#4}  \end{center}

}   %%% end of newcommand regpreprint

\newcommand{\regaddress}{
Institut f\"ur Theoretische Physik, Universit\"at Regensburg,\\
Postfach 10 10 42, W-8400 Regensburg, Germany
}

\regpreprint{\large\bf}{October 19, 1992}{TPR-92-41, AZPH-TH/92-25}
{\large\bf
Three-dimensional statistical field theory\\
for density fluctuations in heavy-ion collisions}

\vspace{10mm}
\begin{center}
Hans C.\ Eggers\footnote{eggers@rphs1.physik.uni-regensburg.de}\\
\regaddress

\mbox{ }\\    \mbox{ }\\
Hans-Thomas Elze\footnote{elze@cernvm.cern.ch}\\
CERN-TH, CH-1211 Geneva 23, Switzerland\\

\mbox{ }\\    \mbox{ }\\
Ina Sarcevic\footnote{ina@ccit.arizona.edu}\\
Department of Physics, University of Arizona, Tucson AZ 85721, USA\\
\mbox{ }
\end{center}

\begin{abstract}
A statistical field theory of particle production is presented
using a gaussian functional in three dimensions. Identifying the
field with the particle density fluctuation results in zero
correlations of order three and higher, while the second order
correlation function is of a Yukawa form. A detailed scheme for
projecting the theoretical three-dimensional correlation onto
data of three and fewer dimensions illustrates how theoretical
predictions are tested against experimental moments in the
different dimensions. An example given in terms of NA35 parameters
should be testable against future NA35 data.
\end{abstract}

\newpage

%%%%%%%%%%%%%%%%%%%%%%%%%%%%%%%%%%%%%%%%%%%%%%%%%%%%%%%%%%%%%%%%%%%%%
\section{Introduction}
%%%%%%%%%%%%%%%%%%%%%%%%%%%%%%%%%%%%%%%%%%%%%%%%%%%%%%%%%%%%%%%%%%%%%

Motivated by the search for a new exotic phase of matter predicted
by QCD as well as a better understanding of nuclear matter at
high energy densities, much experimental and theoretical effort is
being directed towards studying heavy ions at high energies
\cite{QM91}. The fact that tens to hundreds of nucleons collide to
produce several hundred particles makes it impossible to calculate
analytically exclusive final states. Large-scale Monte Carlo
simulations of microscopic processes and statistical methods have
so far proven the only viable approaches to the problem.

In the realm of statistical analyses, the measurement and theoretical
treatment of one-particle observables such as cross sections,
$p_\perp$ distributions etc.\ have, after a period of dormancy,
recently been supplemented by revived interest in correlations.
Different theoretical attempts to describe correlations in various
regions of phase space, classified under headings such as
Bose-Einstein correlations \cite{Bauer}, intermittency
\cite{Ring,BP} and effects of collective flow \cite{Heinz},
have met with some success but much remains unclear. Some recent
experimental work sees the two as connected, saying that
correlations of like-sign particles measured in intermittency
analyses can be understood as a Bose-Einstein effect \cite{EMC-BE},
while conversely analyzing factorial moments in terms of the
correlation integral and the four-momentum transfer $Q^2$ is
equivalent to the Bose-Einstein analysis \cite{CI}; in hadronic
collisions, UA1 has found a power-law dependence in their
Bose-Einstein analysis \cite{UA1}.

While the languages may be converging, this does
not mean that there is as yet a solid theoretical basis for these
correlations, especially those of higher order: it is still
most difficult to calculate measurable correlation functions from
first principles. It is therefore helpful to explore simple
effective field theories which ultimately should be derived
from the underlying theory of QCD. Thus the situation bears some
resemblance to the pre-BCS era in the research of superconductivity.
What we are looking for is a strong interaction analogy to
the phenomenologically successful Ginzburg-Landau theory
\cite{Fetter}.

The present paper attempts to understand correlations in heavy-ion
collisions in terms of a statistical field theory based on a gaussian
approximation of a suitably defined Ginzburg-Landau theory.
This approach was first used by Scalapino and Sugar \cite{Sca73a}
who identified the field as a pion amplitude. In our case, the field
is not an amplitude or particle density but a fluctuation from the
mean density, an ansatz used previously \cite{Dremin,Elze}. It is
motivated, as we shall see, by the fact that it matches experimental
findings that in heavy-ion collisions only second order correlations
are found while higher order correlations are negligible. In contrast
to our previous work which was related to the rapidity variable only,
all quantities are here calculated strictly in the three-dimensional
phase space of rapidity $y$, azimuthal angle $\phi$ and transverse
momentum $p_\perp$ and then projected down for comparison with the
data (see, however, the comment in Section VI concerning the proper
choice of variables).

The paper is organized as follows. In Section II, we remind the
reader of the basic equations for extracting true correlations
and present the experimental evidence for neglecting correlations
of order higher than two. In Section III, we recapitulate and
develop the formalism of statistical field theories in terms of
a functional formulation and derive from a three-dimensional
functional ansatz a general form of the second order correlation.
We show in Section IV how one integrates out variables to find
moments in one, two and three dimensions and how this tests
theoretical correlation functions. Using NA35 parameters and
$p_\perp$ distributions, we show horizontal and vertical moments
in all dimensions in Section V. A summary and discussion of some
important issues conclude the paper in Section VI.

%%%%%%%%%%%%%%%%%%%%%%%%%%%%%%%%%%%%%%%%%%%%%%%%%%%%%%%%%%%%%%%%%%%%%
\section{Cumulants in heavy-ion collisions}
%%%%%%%%%%%%%%%%%%%%%%%%%%%%%%%%%%%%%%%%%%%%%%%%%%%%%%%%%%%%%%%%%%%%%

Traditionally, correlations were measured as a function of the
distance between bins in phase space while keeping the bin sizes
fixed. Following the proposal to look for a power-law structure in
the correlation function \cite{BP}, a commonly used alternative has
become the measurement of normalized factorial moments as a function
of decreasing bin size while disregarding all distance information.

The factorial moments $F_q(M)$ are constructed as follows. (For the
purposes of this paper we shall stick to the coordinates rapidity,
azimuthal angle and transverse momentum, but the formulation is true
for all variables.) A given total interval
$\Omega_{\rm tot} = \Delta Y\, \Delta \phi\, \Delta P$ is
subdivided into $M^3$ bins of side lengths
$(\Delta Y/M,\, \Delta \phi/M,\, \Delta P/M)$. With $n_{klm}$ being
the number of particles in bin $(k,l,m)$ and
$n^{[q]} \equiv n(n-1)\ldots(n-q+1)$,
the ``vertical'' factorial moment is
%%aps \FL
\begin{eqnarray}
\label{eq:cua}
F_q^v(M) &\equiv& {1\over M^3}
\sum_{k,l,m=1}^M
{
\langle n_{klm}^{[q]} \rangle
\over
\langle n_{klm} \rangle^{q}
} \nonumber\\
&=&
{1\over M^3}
\sum_{k,l,m=1}^M
{
\int_{\Omega_{klm}} \prod_i  d^3 x_i \, \rho_q(x_1\ldots x_q)
\over
\left[\int_{\Omega_{klm}} d^3 x \, \rho_1(x)\right]^q
} .
\end{eqnarray}
The second equality illustrates how the factorial moment can be
written in terms of integrals of the correlation function $\rho_q$
($\Omega_{klm}$ is the region of integration over bin $k,l,m$).
Because for small bin sizes $n_{klm}$ becomes small and the relative
error correspondingly large, an alternative definition is often
preferred for three-dimensional analysis, the ``horizontal''
factorial moment,
%%aps \FL
\begin{eqnarray}
\label{eq:cub}
F_q^h(M) &\equiv& {1\over M^3}
\sum_{k,l,m=1}^M
{
\langle n_{klm}^{[q]} \rangle
\over
\left(\langle N \rangle/M^3\right)^{q}
}  \nonumber\\
&=&
M^{3(q-1)}
\sum_{k,l,m=1}^M
{
\int_{\Omega_{klm}} \prod_i  d^3 x_i \, \rho_q(x_1\ldots x_q)
\over
\left[\int_{\Omega_{\rm tot}} d^3 x \, \rho_1(x)\right]^q
} .
\end{eqnarray}
The latter form, while being much more stable, has the serious
drawback that it is influenced by the shape of the one-particle
distribution function $\rho_1$.

When one or more of the arguments of the correlation function become
statistically independent it factorizes into lower order parts.
To measure true particle correlations, known as cumulants, the
trivial background must be subtracted. The first two cumulants
are defined as \cite{Stu87a}
\begin{eqnarray}
\label{eq:cuba}
C_2(x_1,x_2) &=& \rho_2(x_1,x_2) - \rho_1(x_1) \rho_1(x_2) \;, \\
C_3(x_1,x_2,x_3) &=& \rho_3(x_1,x_2,x_3) - \rho_1(x_1)\rho_2(x_2,x_3)
\nonumber \\
&-& \rho_1(x_2)\rho_2(x_1,x_3) - \rho_1(x_3)\rho_2(x_1,x_2)
\nonumber \\
&+& 2 \rho_1(x_1)\rho_1(x_2)\rho_1(x_3) \,.
\end{eqnarray}
By integrating these well-known relations over each bin, one can
derive equations for integrated cumulants
$K_q^v = \int \prod dx_k C_q / (\int dx \rho_1(x))^q$
in terms of the vertical factorial moments of Eq.\ (\ref{eq:cua}):
the first two are
\begin{equation}
\label{eq:cuc}
K_2^v = F_2^v - 1, \ \ \ \ \ \ K_3^v = F_3^v - 3F_2^v + 2 \;,
\end{equation}
and so on for higher orders \cite{Car90d}. Whenever there are no
true correlations, these cumulants become zero.

Analyses of the $K_q$'s is now being carried out routinely in
conjunction with the standard factorial moments as a function of
bin size. Interestingly, it was found that in the case of heavy-ion
collisions, there are only two-particle correlations: while $K_2$
is positive for all bin sizes, the values of $K_3$, $K_4$ and $K_5$
have all been found to be consistent with zero. While initially
this was found in terms of one-dimensional rapidity data,
measurements by NA35 in two and three dimensions have confirmed
this fact \cite{Seyboth}. As an example we show in Fig.\ 1 the
third order cumulant derived from NA35 200 A GeV Oxygen-Gold
measurements of the factorial moments \cite{Kad92a}. (This
data is given in terms of ``flattened variables''.) The higher
orders $K_4$ and $K_5$ were also found to be consistent with zero.
Corresponding findings for other nuclei and energies  were
published before \cite{Elze,Diss,Car91a}.

%%%%%%%%%%%%%%%%%%%%%%%%%%%%%%%%%%%%%%%%%%%%%%%%%%%%%%%%%%%%%%%%%%%%%
\section{Statistical field theory}
%%%%%%%%%%%%%%%%%%%%%%%%%%%%%%%%%%%%%%%%%%%%%%%%%%%%%%%%%%%%%%%%%%%%%

For the large number of particles produced in high-energy heavy-ion
collisions, a statistical theory of particle production is justified.
A fruitful starting point has been an analogy first drawn by Feynman
and Wilson \cite{Wil73a}, who pointed out that, by interpreting
$(y,\phi,p_\perp)$ as a ``spatial'' coordinate $\vec x$, the final
$N$-particle phase space could be treated as if it were a gas,
bounded by ``walls'' made up of the overall kinematic constraints.
This analogy then carries over into a correspondence of the total
cross section to a (grand canonical) partition function of a gas,
while the $n$-particle cross section becomes the $n$-particle
distribution function of the gas. It immediately leads to an
identification of the probability of producing a secondary with a
certain momentum $(y,\phi,p_\perp)$ with the gas ``density''
$\rho_1$ at the corresponding point $\vec x$.

Botke, Scalapino and Sugar \cite{Sca73a,Bot74} utilized these ideas
to write down a model for particle production. In analogy to
ordinary statistical mechanics, where the density matrix governing
the weights of states can be written in terms of the free energy,
$\hat\rho \propto \exp(-\beta(\hat H-\Omega))$, they defined a
functional $F[\Pi]$ of a random field $\Pi$, which governs the
number of particles produced at point $\vec x$ via
\begin{equation}
\label{eq:sfb}
{1\over \sigma}{d^3\sigma \over d^3\vec x}
= \langle \Pi^2(\vec x) \rangle
= {1\over Z} \int {\cal D}\Pi\, e^{-F[\Pi]} \Pi^2(\vec x) \;,
\end{equation}
where one identifies the particle density with $\Pi^2$ and
$Z \equiv \int {\cal D}\Pi\, \exp(-F[\Pi])$ plays the role of the
partition function. Higher order correlation functions can easily
be found by the appropriate functional derivatives.

The essence of this ansatz is that all the physics is hidden in the
form of the functional $F[\Pi]$ which reduces the greatly redundant
information contained in the many microscopic degrees of freedom to
just a few phenomenological parameters. (Botke, Scalapino and Sugar
showed that  their classical functional formalism is isomorphic to
a system of quantized Boson fields with normal ordering
\cite{Bot74}.)

The analogy with the free energy permits the utilization of the
Ginzburg-Landau expansion of the free energy near a second-order
phase transition \cite{Lan}, which uses functionals of the type
\begin{equation}
\label{eq:sfc}
F[\Pi(\vec x)] = \int d^3 x \left[ \alpha (\nabla_x \Pi)^2 +
            \mu^2 \Pi^2 + \lambda (\Pi^2)^2 \right] ,
\end{equation}
giving a minimum expectation value of the order parameter at  zero
above and nonzero below the phase transition. Applied to low-energy
hadronic data by Scalapino and Sugar, it has recently been used for
KNO scaling \cite{Car87b} and in an attempt to find a critical
index for a phase transition to a quark-gluon plasma \cite{Hwa92}.
Indeed, recent lattice gauge calculations suggest that, while a
first-order transition is likely for the pure glue SU(3) gauge
theory and for more than three quark flavors, there is increasing
likelihood of a second-order phase transition for two quark flavors
\cite{Lat}, and Wilczek \cite{Wil92a} has made first attempts at
formulating such a transition in terms of a chiral order parameter.

However, our aim here is more modest in that we do not specify the
dependence of the constants $\alpha$, $\mu$ and $\lambda$ entering
the free energy functional in Eq.\ (\ref{eq:sfc}) on the dynamical
parameters of the heavy-ion reaction under consideration such as
total center-of-mass energy, mass numbers and impact parameter.
Instead, we treat them as phenomenological parameters to be
determined from the experimental data and to be interpreted later
on theoretically with the help of a more microscopic model as
discussed in Section VI. Also we do not consider the definition of
the functional of Eq.\ (\ref{eq:sfd}) below an expansion valid for
small values of the field $\Phi$ only, and we cannot draw any
conclusions whether the system undergoes a phase transition or not.
An exception would arise when drastic and systematic variations in
the phenomenological parameters as deduced from one experiment
compared to another are observed (see Section V).

We define a random field $\Phi$ as a function in the
three-dimensional space spanned by $(y,\phi,p_\perp)$.
Throughout, $p_\perp$ will be implicitly divided by a constant
scale $\cal P$ so that it is dimensionless. Since we are not
looking for a phase transition, we omit the quartic term,
$\lambda \approx 0$, and start with the functional \cite{Elze}
%%aps \widetext
\begin{equation}
\label{eq:sfd}
F[\Phi] = \int_0^P dy\, \int_{-P/2}^{P/2} d^2p_\perp
\left[ a^2 \left(\partial\Phi \over \partial y\right)^2
     + a^2 \left(\nabla_{\vec p_\perp} \Phi \right)^2
     + \mu^2 \Phi^2
\right] \;.
\end{equation}
%%aps \narrowtext
We note that $F$ is not rotationally invariant but rather
boost-invariant in the direction of the collision axis, i.e.\
tailored to the specific symmetry of the collision. The
integration bound $P$ is chosen the same for $y$ and $p_\perp$
for simplicity but does not have to be the same in general.

The expectation value of any function of $\Phi$ is found by taking
functional derivatives with respect to a source term, $J\cdot \Phi$,
added into the integrand of $F$ in Eq.\ (\ref{eq:sfd}) and thus into
the exponential entering the functional integral for the partition
function
$Z$  (take $\vec x_i \equiv (y_i,\phi_i,p_{\perp i})$),
\begin{eqnarray}
\label{eq:sfe}
\langle \Phi(\vec x_1)\ldots \Phi(\vec x_k) \rangle
&=&
{1\over Z} \int {\cal D}\Phi e^{-F[\Phi]}\, \Phi(\vec x_1)\ldots
         \Phi(\vec x_k) \nonumber \\
&=&
{1\over Z}
{\delta^k Z[F,J] \over \delta J(\vec x_1)\ldots \delta J(\vec x_k)
}\biggr\vert_{J = 0} \;.
\end{eqnarray}
For the functional (\ref{eq:sfd}), we find the three-dimensional
form of the two-point function
\begin{equation}
\label{eq:sff}
\langle \Phi(\vec x_1)\Phi(\vec x_2) \rangle
= {1\over 8 \pi a^2} {e^{-R/\xi} \over R} \;,
\end{equation}
where $\xi = a/\mu$, and with $p_i \equiv p_{\perp i}$,
%%aps \FL
\begin{equation}
\label{eq:sfg}
R  \equiv
\sqrt{(y_1-y_2)^2 + p_1^2 + p_2^2 - 2p_1 p_2\cos(\phi_1-\phi_2) }
\;.
\end{equation}
So far, we have not defined the field $\Phi$ in terms of physical
observables. For reasons that will become apparent shortly, we
define $\Phi(\vec x)$ as the fluctuation at the point $\vec x$
of the particle density for a given event, $\hat\rho_1(\vec x)$,
above/below the mean single particle distribution $\rho_1$ at that
point:
\begin{equation}
\label{eq:sfh}
\Phi(\vec x) \equiv {\hat\rho_1(\vec x) \over \rho_1(\vec x) } - 1 \;.
\end{equation}
Experimentally, $\Phi$ is the (normalized) difference between
the event histogram and the event-averaged one-particle distribution
for a given bin.

This definition was previously used by Dremin and Nazirov
\cite{Dremin} and in a one-dimensional context by Elze and Sarcevic
\cite{Elze}. The rationale in the present and previous case is the
same: all cumulants except the second order cumulant become exacly
zero. To see how this comes about, we note that
$\rho_q(\vec x_1,\ldots,\vec x_q) =
\langle \hat\rho_1(\vec x_1)\ldots\hat\rho_1(\vec x_q)\rangle$,
and with the definition (\ref{eq:sfh}) and the relations between
the $\rho_q$ and cumulants $C_q$ one finds that the expectation
values of $\Phi$ can be written in terms of the reduced cumulants
$k_q(\vec x_1,\ldots,\vec x_q)
\equiv C_q(\vec x_1,\ldots,\vec x_q)/
       \rho_1(\vec x_1)\ldots\rho_1(\vec x_q)$
as
%%aps \FL
\begin{eqnarray}
\label{eq:sfi}
\langle \Phi(\vec x_1)\Phi(\vec x_2) \rangle \;
&=& \;  k_2(\vec x_1,\vec x_2)\;,
\\
\langle \Phi(\vec x_1)\Phi(\vec x_2)\Phi(\vec x_3) \rangle \;
&=&\;  k_3(\vec x_1,\vec x_2,\vec x_3)   \;, \\
\langle  \Phi(\vec x_1)\Phi(\vec x_2)\Phi(\vec x_3)\Phi(\vec x_4)
\rangle \;
&=&\; k_4(\vec x_1,\vec x_2,\vec x_3,\vec x_4) \\
&+&
\sum_{(3)} k_2(\vec x_1,\vec x_2)k_2(\vec x_3,\vec x_4)
\nonumber  \;,
\end{eqnarray}
the sum running over 3 permutations in the arguments. On the
other hand, the form of the functional (\ref{eq:sfd}) ensures that
all expectation values of an odd number of $\Phi$'s are zero,
while for the fourth order,
%%aps \FL
\begin{equation}
\label{eq:sfk}
\langle \Phi(\vec x_1)\Phi(\vec x_2)\Phi(\vec x_3)\Phi(\vec x_4)
\rangle =
\sum_{(3)} \langle \Phi(\vec x_1)\Phi(\vec x_2) \rangle
           \langle \Phi(\vec x_3)\Phi(\vec x_4) \rangle ,
\end{equation}
so that $k_4 = 0$ also, and similarly for the higher even cumulants.
These equations are actually just the relations between cumulants
and central moments, see Eq.\ (3.38) in \cite{Stu87a}.

In summary, the specific form of the functional (\ref{eq:sfd}) and
the definition of $\Phi$ as a fluctuation field ensure that all
cumulants of order 3 and higher are zero, while
$k_2(\vec x_1,\vec x_2)$ has the Yukawa form (\ref{eq:sff}).

We conclude this section with some comments:
\begin{enumerate}
\item
Of course with better statistics it may eventually become clear
that there is some small residual cumulant of higher order; the
presently large error bars would permit that. Were this to happen,
the theory in its present form would have to be modified by, for
example, the inclusion of interaction terms of higher order in
$\Phi$, with or without additional implications for the existence
of a phase transition.
\item
The present model, where the higher order cumulants are exactly
zero, is incompatible with the {\it linking ansatz} of Carruthers
and Sarcevic \cite{Car89c} used for {\it hadronic} collisions,
in which higher order cumulants are products of $k_2$ according
to $k_q = A_q k_2^{q-1}$, an ansatz which for certain values of
the constants $A_q$ yields the negative binomial distribution.
So far, there is no direct evidence for linking in heavy-ion
collisions \cite{Diss}; however, the present size of error bars
does not permit any conclusions on this point.
\item
Thirdly, there has been some discussion whether cumulants of higher
order can be identically zero consistently. Since this is a technical
question with no consequences for our further development, we defer
this point to the Appendix.
\end{enumerate}

%%%%%%%%%%%%%%%%%%%%%%%%%%%%%%%%%%%%%%%%%%%%%%%%%%%%%%%%%%%%%%%%%%%%%
\section{Projecting down to lower dimensions}
%%%%%%%%%%%%%%%%%%%%%%%%%%%%%%%%%%%%%%%%%%%%%%%%%%%%%%%%%%%%%%%%%%%%%

Apart from the fact that all higher order cumulants are exactly zero
by construction, the main result of our model is the Yukawa form for
the second reduced cumulant, $k_2 \propto e^{-R/\xi}/R$.
This can be compared to data only after a suitable integration over
its variables. In its present form, $k_2$ is not a power law of any
of its variables. Setting $\cos \theta = 1$, one can show
analytically that
$\int_0^{\delta y}d|y_1-y_2| \int_0^{\delta p}dp_1\,dp_2\, R^{-1}$
becomes linear in $\delta y\delta p$, while numerical integration
over all variables yields
$K_2 \propto (\delta y \delta \phi \delta p)^{-\alpha}$, with
$\alpha = 1$ within error, i.e.\ the integrated version
resembles a power law.

We now derive the detailed form of integrated cumulants $K_2$ for
various dimensions in terms of the three-dimensional
$k_2(\vec x_1,\vec x_2)$. For this purpose, the ``vertically''
and ``horizontally'' normalized versions Eqs.(\ref{eq:cua})
and (\ref{eq:cub}) have to be treated separately.

The second order vertical cumulant is, in three dimensions,
(always taking $\vec x \equiv (y,\phi,p_\perp)$)
%%aps \FL
\begin{equation}
\label{eq:prc}
K_2^v(\delta y,\delta\phi,\delta p)
=  F_2^v - 1
=   {1\over M^3}\sum_{k,l,m=1}^M K_2^v(k,l,m) \;,
\end{equation}
with
%%aps \widetext
\begin{eqnarray}
\label{eq:prd}
K_2^v(k,l,m)    &=&
{       \langle n_{klm}(n_{klm}-1) \rangle
      - \langle n_{klm} \rangle^2
\over   \langle n_{klm} \rangle^2 }
=
{
\int_{\Omega_{klm}}d^3\vec x_1\, d^3 \vec x_2\; C_2(\vec x_1,\vec x_2)
\over
\left[
\int_{\Omega_{klm}}d^3 \vec x\; \rho_1(\vec x)
\right]^2
}  \nonumber\\
&=&
\int_{\Omega_{klm}} d^3\vec x_1\, d^3\vec x_2\; k_2(\vec x_1,\vec x_2)
{
\rho_1(\vec x_1) \rho_1(\vec x_2)
\over
\left[\int_{\Omega_{klm}} d^3 \vec x\, \rho_1(\vec x)
\right]^2
} \;.
\end{eqnarray}
In other words, the integration of $k_2$ to compare with data
involves a correction due to the shape of the one-particle
three-dimensional distribution function $\rho_1(\vec x)$. Equation
(\ref{eq:prd}) as it stands is exact: a knowledge of the theoretical
reduced cumulant $k_2$ can only be translated into a measurable
factorial cumulant $K_2$ when the full three-dimensional
one-particle distribution is known and taken into account (of
course the same is true for
$r_2 \equiv \rho_2/\rho_1\rho_1$ versus $F_2$).

Knowledge of $k_2(\vec x_1,\vec x_2)$ and $\rho_1(\vec x)$ in
three dimensions can also immediately be used to compare to factorial
cumulant data of lower dimensions. For example, in $(y,\phi)$,
the cumulant is
$K_2^v(\delta y,\delta\phi) = M^{-2} \sum_{lm} K_2^v(l,m)$ with
the transverse momentum integrated over the whole window $\Delta P$
({\it cf.} Section VI),
\begin{eqnarray}
\label{eq:prf}
K_2^v(l,m) &=&
{      \langle n_{lm}(n_{lm}-1)\rangle
     - \langle n_{lm} \rangle^2
\over
       \langle n_{lm} \rangle^2 } \nonumber \\
&=&
\int_{\Omega_m}dy_1 dy_2 \int_{\Omega_l}d\phi_1 d\phi_2
\int_{\Delta P} dp_1 dp_2\,
{
 k_2(\vec x_1,\vec x_2)\, \rho_1(\vec x_1)\rho_1(\vec x_2)
\over
\left[ \int_{\Omega_m}dy \int_{\Omega_l}d\phi \int_{\Delta P} dp\;
\rho_1(\vec x) \right]^2
} ,
\end{eqnarray}
and the cumulant for the rapidity only is
$K_2^v(\delta y) = M^{-1}\sum_m K_2^v(m)$, with
$K_2^v(m)$  an integral just like Eq.\ (\ref{eq:prf}) but
with $\int_{\Omega_l}$ being replaced by $\int_{\Delta \phi}$.
Cumulants of other variable combinations are obtained analogously.

For the horizontal normalization, the three-dimensional cumulant
$K_2^h = M^{-3}\sum_{klm}K_2^h(k,l,m)$ consists of
\begin{equation}
\label{eq:prg}
K_2^h(k,l,m) =
{
       \langle n_{klm}(n_{klm}-1) \rangle
     - \langle n_{klm} \rangle^2
\over
       \left(\langle N \rangle_{\Omega}/M^3\right)^2
}   ,
\end{equation}
where $\langle N \rangle_{\Omega}$ is by definition the
number of particles within the experimentally defined total volume
$\Omega_{tot} = \Delta Y \Delta \phi \Delta P$. Care must be taken
to define theoretical quantities such that they are normalized to
this experimental domain. Keeping this in mind, we can write
\begin{equation}
\label{eq:pri}
K_2^h(k,l,m) = M^6
\int_{\Omega_{klm}} d^3\vec x_1\, d^3\vec x_2\; k_2(\vec x_1,\vec x_2)
{
\rho_1(\vec x_1) \rho_1(\vec x_2)
\over
\left[\int_{\Omega_{tot}} d^3 \vec x\, \rho_1(\vec x)
\right]^2
} \;.
\end{equation}
%%aps \narrowtext
Projections onto two dimensions $(y,\phi)$ are then of the form of
Eq.\ (\ref{eq:prf}) but with the prefactor $M^4$ and with
$\Omega_{tot}$ replacing the bin region integrals $\Omega_l$,
$\Omega_m$, $\Delta P$ in the denominator. For one-dimensional
data in $y$, the prefactor is $M^2$ and $\Omega_l$ becomes
$\Delta\phi$ in the numerator and $\Omega_{tot}$ in the
denominator integrals. Corresponding versions can be written down
for $K_2^h(\delta\phi)$ and $K_2^h(\delta p)$.

For horizontal factorial moments, it is important to remember that
the simple relations between factorial moments and cumulants,
Eq.\ (\ref{eq:cuc}), are not valid but that rather
\begin{equation}
\label{eq:hork}
F_2^h = K_2^h + \sum_{k,l,m = 1}^M \langle n_{klm} \rangle^2
/ \langle N \rangle_{\Omega}^2
\end{equation}
for three dimensions, with corresponding relations for lower
dimensions. This has to be taken into account in an eventual
comparison with data.

With these relations it is thus possible, given any
three-dimensional theoretical function $k_2$ (or $r_2$), to
compute factorial cumulants and moments for any combination of
its variables. Doing this for different variables can serve as
a strong test of the theoretical function as the moments probe
its different regions.

%%%%%%%%%%%%%%%%%%%%%%%%%%%%%%%%%%%%%%%%%%%%%%%%%%%%%%%%%%%%%%%%%%%%%
\section{Comparison with the data}
%%%%%%%%%%%%%%%%%%%%%%%%%%%%%%%%%%%%%%%%%%%%%%%%%%%%%%%%%%%%%%%%%%%%%

So far, there are only two published sets of heavy-ion factorial
moment data spanning several dimensions: one for 200 A GeV
Sulfur-Gold collisions measured by the EMU01 collaboration
\cite{Dublin,EMU01-90a} and  one- and two-dimensional data of the
KLM collaboration \cite{KLM}. Both are unfortunately not suitable
for our analysis, the former because of probable contamination of
the $(\eta,\phi)$ moment by gamma conversion \cite{LUIP}, the
latter because the data is binned with different $M$ values in
each variable. Both, being emulsion experiments, can measure only
up to two dimensions. (An example of EMU01 analysis can be found
in Ref.\ \cite{HCE}.)

An analysis by NA35 of 200 A GeV Oxygen-Gold moments in all
dimensions is in preparation but not yet available; also,
prospective Sulfur-Sulfur moments will provide further tests of
the model \cite{Ivopc}. In anticipation of such data, we have made
a detailed study of moments for NA35 experimental parameters and
their measured O+Au transverse momentum distribution. Lacking the
appropriate data, we have set the parameters to arbitrary but
plausible values $a=2.0$, $\xi = 1.0$, hoping that exact fits can
be made when experimental moments become available.

Since all the equations of the previous section are exact,
they can be used in their unabridged forms whenever the full
three-dimensional $\rho_1(\vec x)$ is known and enough CPU time
is available. In making comparisons with data, it will however
usually be necessary to approximate these exact forms, partly
because three-dimensional one-particle distributions are not
available, partly to save on computer time. We have therefore
made the following approximations in our simulation:

1.
We factorize the three-dimensional one-particle distribution into
its separate variables:
\begin{equation}
\label{eq:dac}
\rho_1(\vec x) = \langle N \rangle_\Omega
                  g(y)\, h(\phi)\, f(p_\perp)\;,
\end{equation}
where the factor $\langle N \rangle_\Omega$ ensures that
the three distributions $g$, $h$ and $f$ are separately normalized
over their respective total intervals $\Delta Y$, $\Delta \Phi$
and $\Delta P$. (This factorization is known not to be true for
some cases \cite{Schu91}.)

2.
The azimuthal distribution is taken as flat, $h(\phi) = 1/\Delta\Phi$.

3.
We use the full experimental parametrization for $f(p_\perp)$
provided by NA35 \cite{NA35-88a};

4.
For the rapidity distribution, we use a gaussian parametrization
with $\sigma = 1.32$ \cite{NA35-88b}, and transform
$g(y_1)g(y_2)$ in the numerator with
$Y = (y_1+y_2)/2$, $y = y_2 - y_1$, to
\begin{equation}
\label{eq:dad}
g(y_1)g(y_2) = {1\over 2\pi\sigma^2}
               e^{-Y^2/\sigma^2}e^{-y^2/4\sigma^2} \;,
\end{equation}
and since $k_2(\vec x_1,\vec x_2)$
is a function of $|y_1-y_2|$ only, we have
$R(\Delta Y) \equiv \sum_m \int_{\Omega_m} dY\, \exp(-Y^2/\sigma^2)
= \int_{\Delta Y} dY\, \exp(-Y^2/\sigma^2)$ as a constant
prefactor, while the remaining gaussian in $y$ is included in the
numerical integration.

5.
As $k_2$ depends on the differences $|y_1-y_2|$ and
$|\phi_1-\phi_2|$ only, we can transform to relative coordinates
and integrate out the center-of-mass coordinate in these variables
(the ``strip approximation'' \cite{Car89c}).

With these approximations, we can derive the complete behavior of
all cumulant moments both for the vertical and horizontal
normalizations. In Fig.\ 2, we show the vertical cumulants for
three-dimensional binning in $(\delta y, \delta\phi, \delta p)$,
for the two-dimensional $(\delta y,\delta\phi)$ and all three
one-dimensional cumulants
$K_2(\delta y)$, $K_2(\delta\phi)$ and $K_2(\delta p_\perp)$.
The pseudo-power-law behavior is clearly
visible for the three-dimensional case, while $K_2$ for two and one
dimension show the familiar saturation caused by the projection
process \cite{Bia90a}. Fig.\ 3 shows the corresponding horizontal
cumulants $K_2^h$ for the same set of variables. All curves of
Figs.\ 2 and 3 are for $a=2.0$, $\xi=1.0$.

A comparison of the two Figures shows that there is virtually no
effect of the nonflat rapidity distribution, while the steep
$p_\perp$ distribution has a large effect in raising both
$K_2^h(\delta y, \delta \phi,\delta p)$ and
$K_2^h(\delta p)$ above their vertical counterparts.
All cumulants must of course have the same value for $M=1$,
independent of their dimension, variable or normalization.
This is also clearly illustrated in the Figures.

Comparing to ``real'' experimental data, the procedure would be
as follows: first, one takes the $F_2$ data of highest dimension
and finds from these the corresponding $K_2$ from Eqs.(\ref{eq:cuc})
or (\ref{eq:hork}), depending on the normalization used.
To this the free parameters $a$ and $\xi$ are fitted using the
appropriate formula of Section III. If the fit is disastrous,
that of course is the end of the game. If it is reasonably good,
further tests of the proposed cumulant $k_2(\vec x_1, \vec x_2)$
are performed by plotting on experimental $K_2$'s of lower
dimension and different variables the relevant theoretical
formulae of Section III keeping $a$ and $\xi$ fixed to their
higher-dimensional best fit values. Of course, one can start out
with lower-dimension best fits and find the corresponding curves
for higher dimensions. This procedure is a nontrivial test of the
theory because, for example, $K_2(\delta y)$ integrates over the
entire regions $\Delta \Phi$ and $\Delta P$ and hence probes the
long-distance behavior of the theoretical function $k_2$, while
the three-dimensional experimental $K_2$ tests only the short-range
behavior of the theoretical $k_2$.

%%%%%%%%%%%%%%%%%%%%%%%%%%%%%%%%%%%%%%%%%%%%%%%%%%%%%%%%%%%%%%%%%%%%%
\section{Conclusions}
%%%%%%%%%%%%%%%%%%%%%%%%%%%%%%%%%%%%%%%%%%%%%%%%%%%%%%%%%%%%%%%%%%%%%

To summarize our work, we studied integrated two-particle and higher
order correlations in the form of factorial moments and cumulants
and made predictions as to the general behavior of horizontal and
vertical moments for the NA35 O+Au data. An exact comparison will
depend on a fit of two parameters, after which everything else
is fixed for all dimensions. It will be interesting to see how
well our model will do with upcoming data from NA35 and hopefully
other experiments.

In particular, we applied a simple statistical field theory model
with a gaussian ``free energy'' functional motivated by analogy
to the Ginzburg-Landau theory of superconductivity, but without
explicitly implying a (second order) phase transition.
Our model, which is defined in terms of the field of density
fluctuations, was first advocated in Ref.\ \cite{Elze} to describe
one-dimensional rapidity correlations as observed in high-energy
heavy-ion collisions. Here, we extended the study to cover one-,
two-, and  three-dimensional correlations as measured through the
intermittency analysis in the full phase space spanned by the
variables rapidity, azimuthal angle, and transverse momentum of
secondary particles.

In our present formulation, the intrinsic scale $\cal P$ for the
momentum cannot be deduced from the integrated dimensionless
moments being compared with experiment; only direct measurements
of the corresponding correlation functions, e.g. $k_q$, would give
enough detailed information for this purpose. Thus, also the mass
parameter and correlation length of our model can only be
determined as dimensionless constants.

It should be stressed that (by construction) our model yields
vanishing cumulant correlation functions of higher order by
construction, $k_{q\ge 3}=0$, which agrees with all presently
available heavy-ion data ({\it cf.} Section II).

Working towards a more microscopic foundation of the
Ginzburg-Landau type model, a comment about the right choice
of kinematic variables seems appropriate: Since any attempt at
deriving the effective three-dimensional statistical field theory
from a more fundamental theory necessarily begins with a
four-dimensional space-time formulation, i.e. leaving the
Feynman-Wilson gas analogy \cite{Wil73a} behind, three- or
four-momenta conjugate to space-time coordinates are the natural
variables. We therefore urge experimentalists to present their
``intermittency'' or correlation data in terms of three-momenta
and with particle identification whenever this is feasible.
Assuming azimuthal symmetry as before, the {\it relevant variables}
for a more refined theoretical analysis seem to be $p_\parallel$,
$p_\perp$, $\phi$.

Of course, the difficult problem how to incorporate the effects
of the very asymmetric  {\it initial conditions} in $p_\parallel$
and $p_\perp$ into the theoretical description remains as
disturbing as before. Presently we attempted to include all
available experimental information here by properly folding in
at least the relevant one-particle distributions into the
projection integrals for the lower-dimensional cumulant moments.
An important consequence of this procedure is that, {\it if} our
model fits a three-dimensional data set, then the projections onto
two- and one-dimensional ones are predicted with no further freedom.

It is also important to point out that functionals defined in
different dimensions produce different results: $K_2$ calculated
from a one-dimensional functional does not correspond to $K_2$
obtained from deriving $k_2$ from a three-dimensional functional
and then projecting down onto one dimension. Thus the formula
obtained analytically in Ref.\ \cite{Elze} from the one-dimensional
functional,
$K_2=2\gamma \xi^2 [(\delta y/\xi)-1+e^{-\delta y/\xi}]/\delta y^2$,
does not hold in our present formulation (even though it may fit
the data).

Concerning the application of a statistical approach to energies
higher than in current heavy-ion collisions, we have to deal with
the expected increasing importance of hard partonic scattering
events (minijets) \cite{QM91}. They will help to populate more
and more the high-$p_\perp$ tails of the one-particle
distributions, where a priori we cannot expect our model to
apply. Whereas presently high momenta are effectively cut off by
the $p_\perp$ distributions, this may become a problem at higher
center-of-mass energies. Eventually a momentum cut-off has to be
introduced to limit the analysis to the same region of soft physics
implicitly investigated here. A thorough treatment of surface
effects in phase space, together with the asymmetric initial
conditions \cite{Sca73a}, then becomes mandatory.

Finally, we point out how a Ginzburg-Landau-type model may be based
on a more fundamental field theory. Assuming that the observed
multiparticle correlations and fluctuations arise in the dense
hadronic phase of matter (after a possible hadronization phase
transition), the first step of a derivation for this conservative
scenario consists in choosing a phenomenologically satisfactory
effective field theory of hadronic interactions such as the sigma
model. (This presents, of course, an alternative to the view that
the relevant fluctuations stem from partonic showering processes
which might seem more natural for high-energy $e^+e^-$-collisions.)
Keeping the high excitation energy of the compressed matter in mind,
one tentatively begins with a field theory at finite temperature,
neglecting for simplicity the difficult non-equilibrium aspects of
heavy-ion collisions. Aiming at a three-dimensional Ginzburg-Landau
statistical picture one is led to integrate out the (imaginary)
time-dependent modes from the path-integral representation of the
(sigma model) field theory. This has so far been achieved in a
one-loop approximation for an arbitrary scalar field theory
yielding a temperature-dependent three-dimensional effective
action\footnote{
Preliminary results of this study were reported by one of us
(H.-Th.E.) at the 1992 Workshop on Finite Temperature Field Theory
at ZiF, Bielefeld, Germany.
}.
It represents the infrared limit of the originally chosen theory in
accordance with ideas on dimensional reduction and can be used as a
starting point to calculate the correlation functions studied in the
present paper (see Section III).

Clearly, much more can be done here to include non-perturbative
effects if one wants to investigate, for example, effects of a
phase transition possibly contained in the chosen field theory.
Formally integrating out the fields of the three-dimensional
effective action in favor of the fields squared (``densities''),
one obtains a Ginzburg-Landau ``free energy'' functional with
its parameters, {\it cf.} Eq.\ (\ref{eq:sfc}), given as
temperature-dependent functions of the renormalized coupling
constants of the original field theory.

Having outlined these further steps in the development of our
model, we conclude that the study of correlations and fluctuations
in dense hadronic matter seems a promising approach to further our
understanding of strong interactions. In particular, the
investigation of one-particle observables alone, which are nicely
reproduced by a wide selection of event generators \cite{QM91},
in principle cannot provide a sufficiently detailed knowledge of the
most interesting (and least understood) soft physics aspects of QCD.

%%%%%%%%%%%%%%%%%%%%%%%%%%%%%%%%%%%%%%%%%%%%%%%%%%%%%%%%%%%%%%%%%%%%%

%%aps \acknowledgements
\mbox{ }

{\bf Acknowledgements:}

We thank I.\ Derado, J.\ Grote, J.\ Schukraft, D.\ Skelding,
E.\ Stenlund and R.J.\ Wilkes for supplying us with the data and
helping us to understand it in numerous discussions. We are indebted
to P.\ Carruthers, A.\ K\"uhmichel and P.\ Seyboth for helpful
remarks and discussions. Special thanks to P.\ Lipa and
E.A.\ de Wolf for helpful comments pertaining to the appendix.
HCE thanks the Alexander von Humboldt Foundation for support.
HThE gratefully acknowledges support by the Heisenberg program
of the Deutsche Forschungsgemeinschaft. This work was supported in
part by the Department of Energy,
Contract No. DE-F602-88ER40456 and DE-F602-85ER40213.

%%%%%%%%%%%%%%%%%%%%%%%%%%%%%%%%%%%%%%%%%%%%%%%%%%%%%%%%%%%%%%%%%%%%%

\newpage

%%aps \begin{references}

%% \newpage

%%%%%%%%%%%%%%%%%%%%%%%%%%%%%%%%%%%%%%%%%%%%%%%%%%%%%%%%%%%%%%%%%%%%%
%%aps \unletteredappendix{\\ Finding multiplicities from cumulants}
\section*{Appendix:\\ Finding multiplicities from cumulants}
%%%%%%%%%%%%%%%%%%%%%%%%%%%%%%%%%%%%%%%%%%%%%%%%%%%%%%%%%%%%%%%%%%%%%

When all factorial cumulants are known, it is usually possible to
derive from them the multiplicity distribution $P_n$. This is done
via the factorial moment generating function
\begin{equation}
\label{eq:apa}
Q(\lambda) = \sum_n (1-\lambda)^n P_n \;,
\end{equation}
which can also be expanded in terms of the unnormalized cumulants
$f_q \equiv \int C_q$ as
\begin{equation}
\label{eq:apb}
Q(\lambda) = \exp\left[ \sum_q {(-\lambda)^q \over q!} f_q  \right]
\,.
\end{equation}
For our model, only the first two cumulants are nonzero,
$f_1 = \bar n$, the average total multiplicity, and
$f_2 = K_2 {\bar n}^2$. With appropriate transformations and the
identity
$P_n = (1/n!)(-\partial/\partial\lambda)^nQ(\lambda)|_{\lambda = 1}$,
one can derive the multiplicity distribution in terms of
Hermite polynomials \cite{Mue71a}
\begin{equation}
\label{eq:apc}
P_n = {i^n\over n!} (f_2/2)^{n/2} e^{-f_1+f_2/2}\;
       H_n\left[ i(f_2-f_1)/\sqrt{2 f_2} \right] ,
\end{equation}
from which, for example, $P_1 = (f_1-f_2)P_0$. Clearly $P_1$
becomes negative for $f_2>f_1$, rendering the multiplicity
distribution invalid. In terms of normalized factorial moments,
this requires $K_2 < {\bar n}^{-1}$ for Eq.\ (\ref{eq:apc})
to be valid. All odd $P_n$'s are similarly dependent on the
sign of the factor $(f_1-f_2)$, while the even ones are
positive.  For most heavy-ion data, the average
multiplicity is very large, meaning that the multiplicity
distribution (\ref{eq:apc}) cannot be trusted for such cases.

Clearly, something is amiss. The resolution of this dilemma
is found on returning to the identification of the normalized
cumulant functions $k_q$ with the expectation values of the
fluctuation fields, Eqs.(\ref{eq:sfi}){\it ff.}
In this identification, the difference between factorial
cumulants and ordinary cumulants is neglected, which for the
large multiplicities of heavy ion collisions is negligible.
Mathematically, however, there is a difference, and this shows
up in the above dilemma for the derivation of $P_n$ from
the factorial cumulants.

If we work with ordinary cumulants, however, we can
unambiguously derive the multiplicity distribution as follows:
the characteristic function \cite{Stu87a}
\begin{equation}
\label{eq:apd}
\phi(t) = \sum_{n=0}^\infty e^{itn} P_n
\end{equation}
is also the sum of ordinary cumulants $\kappa_q$
(e.g. $\kappa_1 = \langle n   \rangle = \bar n$;
$\kappa_2 =       \langle n^2 \rangle
                - \langle n   \rangle^2$),
\begin{equation}
\label{eq:ape}
\phi(t) = \exp\left[
            \sum_{q=1}^\infty {(it)^q \kappa_q \over q! }\right] ,
\end{equation}
which for our truncated cumulant set is
\begin{equation}
\label{eq:apf}
\phi(t) = \exp\left[it{\bar n} - t^2 \kappa_2 / 2 \right]
\end{equation}
and since (up to a normalization constant) the multiplicity
distribution is
\begin{equation}
\label{eq:apg}
P_n = \int_{-\infty}^\infty e^{-itn} \phi(t)\; dt \;,
\end{equation}
we get
\begin{equation}
\label{eq:aph}
P_n \propto \exp\left[ -(n - \bar n)^2 / 2 \kappa_2 \right] ,
\end{equation}
just an ordinary gaussian multiplicity distribution which is
defined perfectly well.

%%%%%%%%%%%%%%%%%%%%%%%%%%%%%%%%%%%%%%%%%%%%%%%%%%%%%%%%%%%%%%%%%%%%%
%%                         FIGURES
%%%%%%%%%%%%%%%%%%%%%%%%%%%%%%%%%%%%%%%%%%%%%%%%%%%%%%%%%%%%%%%%%%%%%

\vspace{4cm}
\noindent
{\Large\bf Figure captions:}

%%aps \figure{
\begin{figure}[h]
\vspace{1cm}
\caption{
Third order cumulant $K_3$ as a function of the number
of bins $M$ for NA35 OAu data in $(y,\phi,p_\perp)$
\protect\cite{Kad92a}.
Cumulants of higher order are also compatible with zero. This fact is
confirmed in analyses in terms of other variables and different
colliding nuclei.}
\end{figure}

%%aps \figure{
\begin{figure}[h]
\vspace{1cm}
\caption{
One function $k_2(\vec x_1,\vec x_2)$ determines all:
Theoretical vertical cumulant moments $K_2^v$ for
various dimensions, for fixed parameters $a=2.0$, $\xi = 1.0$,
incorporating the experimental NA35 rapidity and $p_\perp$
distributions for 200 A GeV O+Au.}
\end{figure}

%%aps \figure{
\begin{figure}[h]
\vspace{1cm}
\caption{
Theoretical horizontal cumulant moments
$K_2^h$ for the same fixed parameters and NA35 distributions
as in Fig.\ 3. The effect of the $p_\perp$ distribution is clearly
visible. Comparison with NA35 data requires
conversion to horizontal factorial moments $F_2^h$ and a fit
of $a$ and $\xi$.}
\end{figure}


\begin{thebibliography}{99}

\bibitem{QM91}Quark Matter '91, Gatlinburg, Tennessee,
             November 11--15, 1991, Nucl.\ Phys.\ {\bf A544} (1992).

\bibitem{Bauer}W.\ Bauer, C.K.\ Gelbke and S.\ Pratt,
               preprint MSUCL-824 (1992), (to be published).

\bibitem{Ring}{\it Proceedings of the Ringberg Workshop on
                Multiparticle Production}, 1991,
               edited by R.C.\ Hwa and W.\ Ochs and N.\ Schmitz
                (World Scientific, 1992).

\bibitem{BP}A.\ Bia\l as and R.\ Peschanski,
            Nucl.\ Phys.\ {\bf B273},703 (1986);
            Nucl.\ Phys.\ {\bf B308},857 (1988).

\bibitem{Heinz}U.\ Mayer, E.\ Schnedermann and U.\ Heinz,
preprint TPR-92-13 (to be published).

\bibitem{EMC-BE}I.\ Derado, G.\ Jancso and N.\ Schmitz,
              preprint MPI-PhE/92-07 (1992), (to be published).

\bibitem{CI}
P.\ Lipa, P.\ Carruthers, H.C.\ Eggers and B.\ Buschbeck,
Phy.\ Lett.\ {\bf 285B},300 (1992);
 H.C.\ Eggers, P.\ Lipa, P.\ Carruthers and B.\ Buschbeck,
(submitted to Phys.\ Rev.\ D).

\bibitem{UA1}UA1 Minimum Bias Collaboration, B.\ Buschbeck
        {\it et al.}, in Proceedings of the XXII International
        Symposium on Multiparticle Dynamics, Santiago de Compostela,
        1992, edited by C.\ Pajares, (to be published).

\bibitem{Fetter}A.L.\ Fetter and J.D.\ Walecka,
            {\it Quantum theory of many particle systems},
            McGraw Hill (1971).

\bibitem{Sca73a}D.\ J.\ Scalapino and R.\ L.\ Sugar,
               Phys.\ Rev.\ D {\bf 8}, 2284 (1973).


\bibitem{Dremin}I.\ Dremin and M.T.\ Nazirov, in \cite{Ring}.

\bibitem{Elze}H.-Th.\ Elze and I.\ Sarcevic,
              Phys.\ Rev.\ Lett.\  {\bf 68}, 1988 (1992).

\bibitem{Stu87a}A.\ Stuart and J.K.\ Ord,
               {\it Kendall's Advanced Theory of Statistics}, Vol.1,
              5th edition, (Oxford University Press, New York 1987).

\bibitem{Car90d}P.\ Carruthers, H.C.\ Eggers, and I.\ Sarcevic,
                Phys.\ Lett. {\bf 254B},258 (1991).


\bibitem{Seyboth}NA35 Collaboration, I.\ Derado, in \cite{Ring};
                P.\ Seyboth, in \cite{QM91}, p.\ 293C.

\bibitem{Kad92a}K.\ Kadija and P.\ Seyboth, Phys.\ Lett.\
              {\bf 287B},363 (1992).

\bibitem{Diss}H.C.\ Eggers, Ph.D.\ thesis,
              University of Arizona (1991).

\bibitem{Car91a}P.\ Carruthers, H.C.\ Eggers and I.\ Sarcevic,
                Phys.\ Rev.\ C {\bf 44}, 1629 (1991).

\bibitem{Wil73a}K.G.\ Wilson, Cornell preprint CLNS-131 (1970),
              published in {\it Proceedings of the 14th Scottish
              Universities Summer School in Physics}, 1973,
              edited by R.\ L.\ Crawford and R.\ Jennings,
              (Academic Press, 1974); R.P.\ Feynman, unpublished.

\bibitem{Bot74}J.C.\ Botke, D.J.\ Scalapino and R.L.\ Sugar,
            Phys.\ Rev.\ D {\bf 9},813 (1974);
            Phys.\ Rev.\ D {\bf 10},1604 (1974).

\bibitem{Lan}L.D.\ Landau and E.M.\ Lifschitz,
             {\it Statistische Physik}, 6th edition,
             (Akademie Verlag, Berlin, 1984).

\bibitem{Car87b}P.\ Carruthers and I.\ Sarcevic,
               Phys. Lett. {\bf 189B},442 (1987).


\bibitem{Hwa92}R.C.\ Hwa and M.T.\ Nazirov, preprint OITS-490 (1992);
               R.C.\ Hwa and J.\ Pan, preprint OITS-496 (1992),
               (to be published).


\bibitem{Lat}For a review, see D.\ Toussaint, invited talk presented
             at the Int.\ Symp.\ LATTICE-91, Tsukuba, Japan, (1991),
             Arizona preprint AZPH-TH/92-1 (to be published).


\bibitem{Wil92a}F.\ Wilczek, preprint IASSNS-HEP-92/23 (1992),
                to be published in the Proceedings of the IFT
                Conference on Dark Matter, 1992;
                Int.\ J.\ Mod.\ Phys.\ {\bf A7}, 3911 (1992).

\bibitem{Car89c}P.\ Carruthers and I.\ Sarcevic,
               Phys.\ Rev.\ Lett.\ {\bf 63}, 1562 (1989).

\bibitem{Dublin}EMU01 Collaboration, R.\ J.\ Wilkes {\it et al.}, in
                {\it Proceedings of the XXII International Cosmic
                 Ray Conference}, vol.\ 4 p.\ 21,
                (Trinity College, Dublin, 1991).

\bibitem{EMU01-90a}EMU01 Collaboration, M.I.\ Adamovich {\it et al.},
                Phys.\ Rev.\ Lett.\ {\bf 65}, 412 (1990).

\bibitem{KLM}KLM Collaboration, R.\ Holynski {\it et al.},
             Phys.\ Rev.\ Lett.\ {\bf 62}, 733 (1989);
             Phys.\ Rev.\ C {\bf 40}, 2449 (1990).

\bibitem{LUIP}EMU01 Collaboration, M.I.\ Adamovich {\it et al.},
preprints UWSEA-PUB-92-07 and LUIP 9202 (to be published).

\bibitem{HCE}H.C.\ Eggers {\it et al.},
in Proceedings of the XXII International Symposium on Multiparticle
Dynamics, Santiago de Compostela, 1992, edited by C.\ Pajares,
(to be published).

\bibitem{Ivopc}I.\ Derado and A.\ K\"uhmichel, private communication.

\bibitem{Schu91}J.\ Schukraft, preprint CERN-PPE/91-04, in
             Proceedings of the International Workshop on Quark
             Gluon Plasma Signatures, 1990 (to be published).

\bibitem{NA35-88a}NA35 Collaboration, H.\ Str\"obele {\it et al.},
             Z.\ Phys.\ {\bf C38}, 89 (1988).

\bibitem{NA35-88b}NA35 Collaboration, W.\ Heck {\it et al.},
             Z.\ Phys.\ {\bf C38}, 19 (1988).

\bibitem{Bia90a}A. Bia\l as and J. Seixas,
             Phys.\ Lett.\ {\bf 250B}, 161 (1989);
             W.\ Ochs, Phys.\ Lett.\ {\bf 247B}, 101 (1990),

\bibitem{Mue71a}A.H.\ Mueller,
                Phys.\ Rev.\ D {\bf 4},150 (1971).

%%aps \end{references}
\end{thebibliography}
\end{document}